\documentclass[final]{aipproc}

\layoutstyle{6x9}

\begin{document}

\title{Modular Inflation and the Curvaton}


\author{George Lazarides}{address={Physics Division,
School of Technology, Aristotle University of
Thessaloniki,\\ GR-54124 Thessaloniki, Greece}}

\begin{abstract}
Supersymmetric Peccei-Quinn models which provide
a suitable candidate for the curvaton field are
studied. These models also solve the $\mu$
problem, while generating the Peccei-Quinn scale
dynamically. The curvaton is a pseudo
Nambu-Goldstone boson corresponding to an angular
degree of freedom orthogonal to the axion. Its
order parameter increases substantially following
a phase transition during inflation. This results
in a drastic amplification of the curvaton
perturbations. Consequently, these models are
able to accommodate low-scale inflation with
Hubble parameter at the TeV scale such as modular
inflation. We find that modular inflation with
the orthogonal axion as curvaton can indeed
account for the observations for natural values
of the parameters. In particular, the spectral
index can easily be made adequately lower than
unity in accord with the recent data.
\end{abstract}

\maketitle

\section{Introduction}

\par
The precise cosmological observations of the last
decade have established inflation (for a review
on inflation, see e.g. \cite{lectures}) as an
essential extension of the hot big bang model.
However, the case of slow-roll inflation (i.e.
the case with inflaton mass $\ll H_*$, the Hubble
parameter at the time when the cosmological
scales exit the horizon during inflation) suffers
from the fact that, typically, supergravity
(SUGRA) introduces \cite{cllsw,randall}
corrections to the inflaton mass of order $H_*$
during inflation. One way to keep
the inflaton mass under control is to use as
inflaton a pseudo Nambu-Goldstone boson (PNGB)
field, since the flatness of the potential of
such a field is protected by a global
${\rm U}(1)$ symmetry. Such candidates are
\cite{modular} the string axions, which are the
imaginary parts of string moduli fields with the
flatness of their potential lifted only by (soft)
supersymmetry (SUSY) breaking. This results in
inflaton masses $\stackrel{_{<}}{_{\sim }} H_*$.
The resulting modular inflation can be of the
fast-roll type \cite{fastroll}. Fast-roll
inflation lasts only a limited number of
e-foldings, which, however, can be enough to
solve the horizon and flatness problems.

\par
In modular inflation, $H_*\sim 1~{\rm TeV}$ and,
thus, the inflationary energy
scale is much lower than the grand unified theory
(GUT) scale. As a result, the perturbations of
the inflaton are not sufficiently large to
account for the required density perturbations for
explaining the large scale structure in
the universe and the temperature perturbations in
the cosmic microwave background radiation (CMBR).
(For a low-scale inflation model where the
inflaton perturbations are adequate, see
\cite{anupam}.) Thus, a curvaton \cite{enqv} (see
also \cite{earlier}), i.e. another ``light''
field during inflation, is necessary to provide
the observed curvature perturbation. However,
even the curvaton cannot \cite{lythbound}
generically help us to reduce the inflationary
scale to energies low enough for modular inflation.
This is possible only in certain curvaton models
which amplify \cite{amplif,lett} additionally the
curvaton perturbations.

\par
We will construct \cite{pngbcd} a class of SUSY
Peccei-Quinn (PQ) models \cite{pq} which possess
such an amplification mechanism and also solve
naturally the strong $CP$ and $\mu$ problems. We
use as curvaton an angular degree of freedom
orthogonal to the QCD axion, which we will call
orthogonal axion (the radial PQ field was used as
curvaton in \cite{curv}). We study the
characteristics of the scalar potential in this
class of models. We then focus on curvaton
physics and derive a number of important
constraints necessary for the model to be a
successful curvaton model. Finally, we
concentrate on a concrete example of this class
of models and show that it can indeed work for
natural values of parameters.

\section{Modular Inflation}

\par
After gravity mediated soft SUSY breaking, the
potential of the inflaton $s$, which is a
canonically normalized string axion, is
\cite{modular}
\begin{equation}
V(s)=V_{\rm m}-\frac{1}{2}m_s^2s^2+\cdots,
\label{Vinf}
\end{equation}
where $V_{\rm m}\sim (m_{3/2}M_{\rm P})^2$ and
$m_s\sim m_{3/2}$ with \mbox{$m_{3/2}\sim 1~
{\rm TeV}$} and \mbox{$M_{\rm P}\simeq 2.44\times
10^{18}~{\rm GeV}$} being the gravitino mass and
the reduced Planck mass respectively. The
ellipsis in \eqref{Vinf} denotes terms which are
expected to stabilize $V(s)$ at
\mbox{$s\sim M_{\rm P}$}. The inflationary
potential $V_*$ at the time when the cosmological
scales exit the horizon is of intermediate scale:
\begin{equation}
V_*^{\frac{1}{4}}\sim \sqrt{m_{3/2}M_{\rm P}}
\sim 10^{10.5}~{\rm GeV}
\end{equation}
for which \mbox{$H_*\sim m_{3/2}$}.

\par
In this model, inflation can be of the fast-roll
type, where
\begin{equation}
s=s_{\rm i}\exp(F_s\Delta N)\quad{\rm with}\quad
F_s\equiv\frac{3}{2}\left(\sqrt{1+\frac{4c}{9}}-1
\right),\quad
c\equiv\left(\frac{m_s}{H_*}\right)^2\sim 1.
\label{Fs}
\end{equation}
Here, $\Delta N$ is the number of the elapsed
e-foldings and $s_{\rm i}$ the initial value of
the inflaton field $s$. From the above, one can
obtain the inflation potential $N$ e-foldings
before the end of inflation as
\begin{equation}
V(N)\simeq V_{\rm m}\left(1-e^{-2F_sN}\right).
\label{VN}
\end{equation}
Even in the fast-roll case, modular inflation
keeps the Hubble parameter $H$ rather rigid.
Indeed, the slow-roll parameter
\begin{equation}
\epsilon_H\equiv-\frac{\dot H}{H^2}=
\frac{1}{2}c^2\left(\frac{s}{M_{\rm P}}
\right)^2\ll 1,
\label{vareps}
\end{equation}
because \mbox{$c\sim 1$} and
\mbox{$s\ll M_{\rm P}$} during inflation
(the overdot denotes derivative with respect to
the cosmic time $t$).

\par
The initial conditions for the inflaton field
are determined by the quantum fluctuations which
send the field off the top of the potential hill.
Hence, we expect that the initial value for the
inflaton is
\begin{equation}
s_{\rm i}\simeq\frac{H_{\rm m}}{2\pi},\quad
{\rm where}\quad
H_{\rm m}\simeq\frac{\sqrt{V_{\rm m}}}{
\sqrt{3}M_{\rm P}}.
\label{s0}
\end{equation}
Assuming that the final value of $s$ is close to
its vacuum expectation value (VEV)
\mbox{$\langle s\rangle\sim M_{\rm P}$}, we find,
from \eqref{Fs}, that the total number of
e-foldings is
\begin{equation}
N_{\rm tot}\simeq\frac{1}{F_s}
\ln\left(\frac{M_{\rm P}}{m_{3/2}}\right),
\label{Ntot}
\end{equation}
where we took into account that
\mbox{$H_{\rm m} \sim m_{3/2}$}.

\section{Amplification of the curvaton
perturbations}

\par
We consider a PNGB curvaton $\sigma$ whose order
parameter $v=v(t)$ (determined by the values of
the radial fields in the model) takes
\cite{amplif,lett} a different (larger)
expectation value in the vacuum than during
inflation and, in particular, when the
cosmological scales exit the horizon. The
potential for the real canonically normalized
field $\sigma$ is
\begin{equation}
V(\sigma)=(v\tilde{m}_\sigma)^2\left[
1-\cos\left(\frac{\sigma}{v}\right)\right]
\quad\Rightarrow\quad V(|\sigma|\ll v)\simeq
\frac{1}{2}\tilde{m}_\sigma^2\sigma^2,
\label{Vs}
\end{equation}
where \mbox{$\tilde{m}_\sigma=
\tilde{m}_\sigma(v)$} is the variable mass of
$\sigma$. In the vacuum, \mbox{$v=v_0$} and
\mbox{$\tilde{m}_\sigma=m_\sigma$}.

\par
In the curvaton scenario, the curvature
perturbation observed by the cosmic microwave
background explorer (COBE) \cite{cobe},
\mbox{$\zeta\simeq 2\times 10^{-5}$}, is given by
\begin{equation}
\zeta\sim\Omega_{\rm dec}\zeta_\sigma,
\label{zeta}
\end{equation}
where $\zeta_\sigma$ is the partial curvature
perturbation of the curvaton and
\mbox{$\Omega_{\rm dec}$} is the ratio of the
curvaton energy density $\rho_\sigma$ to the
total energy density of the universe $\rho$ at
the time of the decay of the curvaton:
\begin{equation}
10^{-2}\stackrel{_{<}}{_{\sim }}\Omega_{\rm dec}
\equiv\left.\frac{\rho_\sigma}{\rho}
\right|_{\rm dec}\leq 1.
\label{r}
\end{equation}
The lower bound originates (see \cite{luw}) from
the $95\%$ confidence level bound on the possible
non-Gaussian component of the curvature
perturbation from the CMBR data obtained by the
Wilkinson microwave anisotropy probe (WMAP)
satellite \cite{wmap}. The partial curvature
perturbation of the curvaton when the latter
oscillates in a quadratic potential is given
\cite{CD} by
\begin{equation}
\zeta_\sigma\sim
\left.\frac{\delta\sigma}{\sigma}
\right|_{\rm dec}\sim
\left.\frac{\delta\sigma}{\sigma}
\right|_{\rm osc},
\label{zs1}
\end{equation}
where ``osc'' denotes the onset of curvaton
oscillations at a time given by $H_{\rm osc}
\sim m_\sigma$ (we assume that
\mbox{$\tilde{m}_\sigma$} had reached its vacuum
value before the onset of oscillations).

\par
The phase $\theta\equiv\sigma/v$ corresponding to
$\sigma$ remains frozen until the oscillations
begin:
\begin{equation}
\theta_{\rm osc}\simeq\theta_*,\quad
\delta\theta_{\rm osc}\simeq\delta\theta_*,
\label{osc*}
\end{equation}
where star denotes the values of quantities at
the time when the cosmological scales exit the
inflationary horizon and \mbox{$\delta\theta$} is
the perturbation in $\theta$. This implies that
\begin{equation}
\left.\frac{\delta\sigma}{\sigma}
\right|_{\rm osc}=\left.\frac{\delta\theta}
{\theta}\,\right|_{\rm osc}\simeq
\left.\frac{\delta\theta}{\theta}\,\right|_*
=\left.\frac{\delta\sigma}{\sigma}\right|_*.
\label{fractional}
\end{equation}
From the curvaton perturbation during inflation
$\delta\sigma_*=H_*/2\pi$, we then find
\cite{pngbcd} that
\begin{equation}
\delta\sigma_{\rm osc}\simeq
\frac{H_*}{2\pi\varepsilon},\quad{\rm where}
\quad\varepsilon\equiv\frac{v_*}{v_0}\ll 1.
\label{dsosc}
\end{equation}
So, after the end of inflation when $v$ assumes
its vacuum value, {\em the curvaton perturbation
is amplified by a factor} $\varepsilon^{-1}$.
Finally, one can show \cite{pngbcd} that
\begin{equation}
\sigma_{\rm osc}\sim
\frac{H_*\Omega_{\rm dec}}
{\pi\varepsilon\zeta}\quad{\rm and}\quad
\varepsilon\geq\varepsilon_{\rm min}
\equiv\frac{H_*}{2\pi v_0},
\label{sosc}
\end{equation}
where we used the relations
\mbox{$\delta\sigma_*/\sigma_*\leq 1$} and \mbox{
$\sigma_{\rm osc}\stackrel{_{<}}{_{\sim }} v_0$}.

\section{Can we construct PQ models with a
PNGB curvaton?}

\par
In the SUGRA extension of the minimal
supersymmetric standard model (MSSM), there exist
D- and F-flat directions in field space which can
generate intermediate scales
\begin{equation}
M_{\rm I}\sim (m_{3/2}
M_{\rm P}^n)^{\frac{1}{n+1}},
\label{eq:inter}
\end{equation}
where $n$ is a positive integer. It is natural to
identify $M_{\rm I}$ with the symmetry breaking
scale $f_{\rm a}$ of the PQ symmetry
${\rm U}(1)_{\rm PQ}$, such
that a $\mu$ term is generated with $\mu\sim
f^{n+1}_{\rm a}/M_{\rm P}^n\sim m_{3/2}$
\cite{kn}. This would simultaneously resolve the
strong $CP$ and $\mu$ problems of MSSM.

\par
For this, we need a non-renormalizable
superpotential term
\begin{equation}
\lambda P^{n+1}h_1h_2/M_{\rm P}^n,
\label{eq:muterm}
\end{equation}
where $\lambda$ is a dimensionless parameter, $P$
is a standard model (SM) singlet superfield with
$\langle P\rangle\sim M_{\rm I}$ and $h_1$, $h_2$
are the electroweak Higgs doublets. One can show
\cite{pngbcd} that $P$ must necessarily carry a
non-zero PQ charge. As a consequence, $P$ has a
flat potential. To lift the flatness of its
potential and generate an intermediate VEV for
$P\sim M_{\rm I}$, we must introduce
\cite{choi,rsym,thermal} a second SM singlet $Q$
with non-zero PQ charge having a coupling of the
type
\begin{equation}
\xi P^{n+3-k}Q^k/M_{\rm P}^n,
\label{eq:PQcoupling}
\end{equation}
where $\xi$ is a dimensionless parameter and $k$
is a positive integer smaller than $n+3$.

\par
After soft SUSY breaking, the scalar potential
possesses \cite{pngbcd} non-trivial minima at
\begin{equation}
|P|,~|Q|\sim (m_{3/2}
M_{\rm P}^n)^{\frac{1}{n+1}},
\label{PQscale}
\end{equation}
where ${\rm U}(1)_{\rm PQ}$ is spontaneously
broken and $f_{\rm a}$ and $\mu$ are generated
dynamically. The soft masses-squared $m_P^2$,
$m_Q^2\sim m_{3/2}^2$ of $P$, $Q$ can have either
sign, while the coefficient $A$ of the soft
$A$-term corresponding to the coupling in
\eqref{eq:PQcoupling}, which is generally complex
with $\vert A\vert\sim m_{3/2}$, must be large
enough for the non-trivial minima to exist if
$m_P^2,~m_Q^2>0$.

\par
To implement our scenario, we need a valley of
local minima of the potential which has negative
inclination and $|P|\ll |P|_0$ and
$|Q|\ll |Q|_0$, where $|P|_0$ and $|Q|_0$ are the
vacuum values of $|P|$ and $|Q|$ respectively. If
the system slowly rolls down this valley during
inflation, the order parameter $v\ll v_0$ and our
amplification mechanism for the curvaton
perturbations may work. This can be achieved only
if one of the masses-squared $m_P^2$, $m_Q^2$ is
negative. Let us take $m_P^2<0$ and $m_Q^2>0$.
In this case, the scalar potential is
\cite{thermal} unbounded below unless $k=1$. So,
we restrict ourselves to the case $k=1$. One can
show \cite{pngbcd} that the orthogonal axion in
this case acquires a mass of order $m_{3/2}$
during inflation and, thus, does not qualify as
a PNGB curvaton.

\par
The addition of a third SM singlet superfield
$S$, however, with a coupling
\begin{equation}
\xi_q P^{n+3-p-q}Q^pS^q/M_{\rm P}^n,
\label{eq:PQScoupling}
\end{equation}
where $p$, $q$ are non-negative integers with
$p+q\leq n+3$ and $q\geq 3$, can drastically
change \cite{pngbcd} the situation allowing the
implementation of our mechanism. In the next
section, we will present a concrete class of
models of this category.

\section{PQ models with an axion-like curvaton}

\par
We consider a class of extensions of MSSM which
are based on the SM gauge group, but also possess
a global anomalous PQ symmetry
${\rm U}(1)_{\rm PQ}$, a global non-anomalous R
symmetry ${\rm U}(1)_{\rm R}$,
and a discrete $Z_2^P$ symmetry. Note, in passing,
that global continuous symmetries can effectively
arise \cite{laz1} from the discrete symmetry
groups of many compactified string theories (see
e.g. \cite{laz2}). In addition to
the usual MSSM superfields $h_1$, $h_2$ (Higgs
${\rm SU}(2)_{\rm L}$ doublets), $l_i$
(${\rm SU}(2)_{\rm L}$ doublet leptons),
$e^c_i$ (${\rm SU}(2)_{\rm L}$ singlet
charged leptons), $q_i$ (${\rm SU}(2)_{\rm L}$
doublet quarks), and $u^c_i$, $d^c_i$
(${\rm SU}(2)_{\rm L}$ singlet anti-quarks)
with $i=1,~2,~3$ being the family index, the
models contain the SM singlet
superfields $P$, $Q$, and $S$. The charges of
the superfields under ${\rm U}(1)_{\rm PQ}$ and
${\rm U}(1)_{\rm R}$ are
\begin{eqnarray}
{\rm PQ}:~P(-2),~Q(2),~S(0),~h_1,~h_2(n+1),
\nonumber\\
{\rm R}:~P(\frac{n+3}{2}),~Q(\frac{n-1}{2}),
~S(\frac{n+1}{2}),~h_1,~h_2(0)
\label{charges}
\end{eqnarray}
with the ``matter'' (quark and lepton)
superfields having ${\rm PQ}=-(n+1)/2$ and
${\rm R}=(n+1)(n+3)/4$. The integer $n$ is
taken to be of the form
\begin{equation}
n=4l+1,\quad{\rm where}\quad l=0,~1,~2,...,
\label{n4l1}
\end{equation}
for reasons to be explained below. Finally, under
$Z_2^P$, $P$ changes sign. Baryon (and lepton)
number is \cite{nonthtripletdec} automatically
conserved to all orders in perturbation theory as
a consequence of ${\rm U}(1)_{\rm R}$ (and
${\rm U}(1)_{\rm PQ}$). The $Z_2$ subgroup of
${\rm U}(1)_{\rm PQ}$ coincides with the matter
parity symmetry $Z_2^{\rm mp}$, which changes the
sign of all matter superfields.

\par
The most general superpotential compatible with
these symmetries is
\begin{eqnarray}
W &=&
y_{eij}l_ih_1e^c_j+y_{uij}q_ih_2u^c_j+
y_{dij}q_ih_1d^c_j
\nonumber \\
& &+\lambda P^{n+1}h_1h_2/M_{\rm P}^n
+\sum_{k=0}^{(n+3)/4}\lambda_k
S^{n+3-4k}(PQ)^{2k}/M_{\rm P}^n,
\label{W}
\end{eqnarray}
where $y_{eij}$, $y_{uij}$, $y_{dij}$ are the
usual Yukawa coupling constants, $\lambda$,
$\lambda_k$ are complex dimensionless
parameters, and summation over the family
indices is implied.

\subsection{The scalar potential}

\par
The resulting scalar potential for PQ breaking
after soft SUSY breaking is
\begin{equation}
V=|F_P|^2+|F_Q|^2+|F_S|^2+ V_{\rm soft},
\label{V}
\end{equation}
where
\begin{equation}
F_P=\sum_{k=1}^{(n+3)/4}2k\lambda_k
\frac{S^{n+3-4k}(PQ)^{2k-1}Q}{M_{\rm P}^n},
\label{FP}
\end{equation}
\begin{equation}
F_Q=\sum_{k=1}^{(n+3)/4}2k\lambda_k
\frac{S^{n+3-4k}(PQ)^{2k-1}P}{M_{\rm P}^n},
\label{FQ}
\end{equation}
and
\begin{equation}
F_S=\sum_{k=0}^{(n-1)/4}(n+3-4k)\lambda_k
\frac{S^{n+2-4k}(PQ)^{2k}}{M_{\rm P}^n}
\label{FS}
\end{equation}
are the F-terms, and
\begin{equation}
V_{\rm soft}=m_P^2|P|^2+m_Q^2|Q|^2+
m_S^2|S|^2
+\left[A\sum_{k=0}^{(n+3)/4}\lambda_k
\frac{S^{n+3-4k}(PQ)^{2k}}{M_{\rm P}^n}+
{\rm h.c.}\right]
\label{Vsoft}
\end{equation}
the soft SUSY-breaking terms. Here, the soft
SUSY-breaking masses-squared $m_P^2$, $m_Q^2$,
and $m_S^2$ are of the order of the $m_{3/2}^2$
and can, in principle, have either sign. However,
the potential $V$ is \cite{pngbcd} bounded below
only if $m_P^2$, and $m_Q^2$ are positive. For
definiteness, we will take these two soft
masses-squared to be equal, i.e. we will put
$m_P^2=m_Q^2\equiv m^2$. Also, for simplicity, we
assumed universal soft SUSY-breaking
$A$-terms with $|A|\sim m_{3/2}$.

\par
\begin{figure}
\includegraphics[height=.3\textheight]{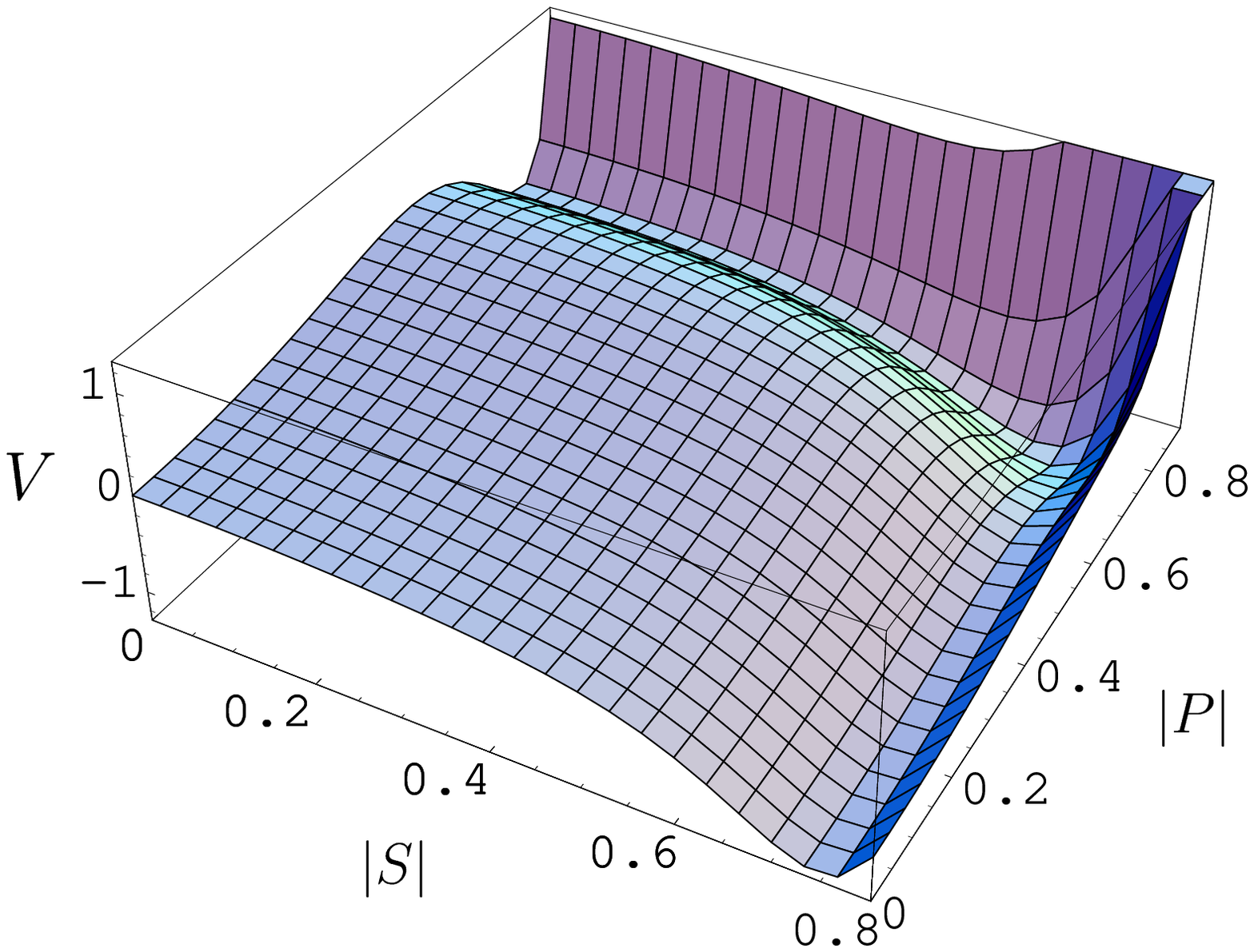}
\caption{Plot of $V$ defined in
\eqref{V}-\eqref{Vsoft} in units of
$M_{\rm I}^{14}/M_{\rm P}^{10}$ with respect to
$|S|$ and $|P|$, which are in units of
$M_{\rm I}$. We took \mbox{$n=5$},
\mbox{$M_{\rm I}\equiv (mM_{\rm P}^5)^{1/6}$},
\mbox{$m_P^2=m_Q^2=-m_S^2\equiv m^2$},
\mbox{$A=-9\,m$}, and \mbox{$\lambda_0=
\lambda_1=\lambda_2=1$}. Also, \mbox{$|P|=|Q|$}
and \mbox{$\theta_S=\theta_P=\theta_Q=0$}
($\theta_S$, $\theta_P$, $\theta_Q$ are the
phases of $S$, $P$, $Q$) so that the potential is
minimized. The trivial and shifted valleys as
well as the trivial and non-trivial minima are
clearly visible.}
\label{fig}
\end{figure}

For reasons which will become clear later,
we take $m_S^2<0$. Therefore, the origin in field
space ($P=Q=S=0$) is a saddle point of the
potential with positive curvature in the $P$ and
$Q$ directions and negative in the $S$ direction.
We will call it trivial saddle point. One can
show \cite{pngbcd} that the potential $V$ has a
``trivial'' valley of local minima which lies on
the $S$ axis (i.e. at $P=Q=0$) and is clearly
visible in Figure~\ref{fig}. Moreover, there
exists a ``trivial'' minimum on this valley at
$|S|\sim (m_{3/2}M_{\rm P}^n)^{1/(n+1)}$,
where ${\rm U}(1)_{\rm PQ}$ is unbroken and no
$\mu$ term is generated. So, we should avoid
ending up at this trivial minimum.

\par
The potential $V$ possesses \cite{pngbcd}
non-trivial minima too with
\begin{equation}
|P|=|Q|,~|S|\sim
(m_{3/2}M_{\rm P}^n)^{\frac{1}{n+1}},
\label{nontrmin}
\end{equation}
where ${\rm U}(1)_{\rm PQ}$ is broken and a
$\mu$ term is generated. Actually, after soft
SUSY breaking (included in $V$), the
only global symmetry surviving in a
non-trivial minimum is $Z_2^{\rm mp}$.

\paragraph{The shifted valley of minima}

We expand $V$ for $|S|\ll |P|\sim |Q|$. The
leading term is
\begin{eqnarray}
V_{(0)}&=&\frac{(n+3)^2}{4}
|\lambda_{\frac{n+3}{4}}|^2
\frac{|PQ|^{n+1}(|P|^2+|Q|^2)}{M_{\rm P}^{2n}}
+m_P^2|P|^2+m_Q^2|Q|^2
\nonumber \\
& &-2|A|
|\lambda_{\frac{n+3}{4}}|
\frac{(|P||Q|)^{\frac{n+3}{2}}}{M_{\rm P}^n}
\cos\left[\frac{n+3}{2}
(\theta_P+\theta_Q)\right],
\label{Vapprox}
\end{eqnarray}
where $\theta_P$ and $\theta_Q$ are the
phases of $P$ and $Q$ respectively and
$A\lambda_{(n+3)/4}$ is taken real and
negative by field rephasing.

\par
The potential $V_{(0)}$ is minimized with
respect to the phases $\theta_P$ and $\theta_Q$
for
\begin{eqnarray}
&\frac{n+3}{2}(\theta_P+\theta_Q)=0\quad
{\rm modulo}\quad 2\pi,\quad|P|=|Q|,&
\nonumber\\
&\left(\frac{|P|^{n+1}}{M_{\rm P}^n}\right)_+
\equiv x_+=\frac{|A|+\sqrt{|A|^2-4(n+2)m^2}}
{(n+2)(n+3)|\lambda_{\frac{n+3}{4}}|}\quad
{\rm for}\quad |A|^2>4(n+2)m^2.&
\label{xpm}
\end{eqnarray}
The presence of the term $\lambda_{(n+3)/4}
(PQ)^{(n+3)/2}/M_{\rm P}^n$ in the superpotential
is vital to the existence of this ``shifted''
minimum. In view of $Z_2^P$, however, this term
can only exist if $(n+3)/2$ is an even positive
integer, which implies the restriction in
\eqref{n4l1}.

\par
For $|S|\ll |P|\sim |Q|$, the shifted minimum of
$V_{(0)}$ is also a minimum of $V$ with respect
to $|P|$ and $|Q|$ at a practically
$S$-independent position. Thus, for small values
of $|S|$, we obtain a ``shifted'' valley of
minima of $V$ at almost constant values of $|P|$
and $|Q|$. This valley, which is clearly visible
in Figure~\ref{fig}, has negative inclination for
non-zero and small values of $|S|$, due to the
negative mass term of $S$. It starts from the
``shifted'' saddle point which lies at $|S|=0$
and $|P|$, $|Q|$ equal to their values at the
shifted minimum of $V_{(0)}$. Note, in passing,
that a shifted valley of minima was first used in
\cite{shift} as an inflationary path in order to
avoid the overproduction of doubly charged
\cite{magg} magnetic monopoles at the end of SUSY
hybrid inflation \cite{cllsw,susyhybrid} in a
Pati-Salam \cite{ps} GUT model.

\paragraph{The PNGB curvaton}

The dominant $S$-dependent part of $V$ can be
expressed as
\begin{eqnarray}
V_{(1)}&=&m_S^2|S|^2-2|A|
|\lambda_{\frac{n-1}{4}}|
\frac{|S|^4(|P||Q|)^{\frac{n-1}{2}}}
{M_{\rm P}^n}\cos\left(4\theta_S+\frac{n-1}{2}
(\theta_P+\theta_Q)\right)
\nonumber \\
& &+\frac{(n-1)(n+3)}{2}|\lambda_{\frac{n-1}{4}}|
|\lambda_{\frac{n+3}{4}}||S|^4(|P|^2+|Q|^2)
\frac{(|P||Q|)^{n-1}}{M_{\rm P}^{2n}}
\nonumber \\
& &\times\cos\left(4\theta_S-2
(\theta_P+\theta_Q)\right),
\label{V1angles}
\end{eqnarray}
where $\theta_S$ is the phase of $S$ and
$A\lambda_{(n-1)/4}$ is made real and negative.
The potential $V_{(1)}$ is minimized with respect
to the phases $\theta_P$, $\theta_Q$, $\theta_S$
for
\mbox{$4\theta_S+(n-1)(\theta_P+\theta_Q)/2=0$}
modulo \mbox{$2\pi$}. Defining the real
canonically normalized fields
\begin{equation}
\phi_P\equiv\sqrt{2}|P|\theta_P,\quad
\phi_Q\equiv\sqrt{2}|Q|\theta_Q,\quad
\phi_S\equiv\sqrt{2}|S|\theta_S,
\label{phiPQS}
\end{equation}
we find \cite{pngbcd} three angular mass
eigenstates on the shifted valley for
$|S|\ll|P|=|Q|$:
\begin{enumerate}
\item
The axion field $a=(\phi_P-\phi_Q)/\sqrt{2}$,
which remains massless to all orders.
\item
$\phi_{PQ}\equiv (\phi_P+\phi_Q)/\sqrt{2}$
with mass-squared $m_{PQ}^2=|A|(n+3)^2
|\lambda_{(n+3)/4}|x_+/2\sim m^2_{3/2}$.
\item
$\phi_S$, our PNGB curvaton (orthogonal axion)
$\sigma$ with suppressed mass-squared
\begin{equation}
\tilde{m}^2_\sigma=\frac{8}{n+2}
|\lambda_{\frac{n-1}{4}}|x_+\frac{|S|^2}{|P|^2}
\Big[(n+5)|A|-(n-1)\sqrt{|A|^2-4(n+2)m^2}\Big]
\sim\frac{|S|^2}{|P|^2}m_{3/2}^2.
\label{massS}
\end{equation}
Note that the $Z_2^P$ symmetry is \cite{pngbcd}
very important for the PNGB nature of $\phi_S$.
\end{enumerate}

\par
SUGRA corrections \cite{cllsw,randall,crisis}
during inflation can be introduced by simply
replacing $A$, $m^2$, and $m^2_S$ by their
effective values:
\begin{equation}
\bar{A}=A+c_AH,\quad\bar{m}^2=m^2+c_{PQ}H^2,
\quad\bar{m}^2_S=m^2_S+c_SH^2=c_SH^2-|m_S^2|
\label{effective}
\end{equation}
respectively. Here, $c_A$ is a complex
parameter of order unity, while $c_{PQ}$ and
$c_S$ are real and positive parameters again
of order unity. We can arrange the parameters so
that $\bar{m}^2_S>0$ during the initial stages
of inflation. In this case, the shifted saddle
point of $V$ becomes a local minimum and the
system may be initially trapped in it. $H$
decreases during inflation and, thus, at some
moment of time, this minimum turns into a saddle
point and the system starts slowly rolling down
the shifted valley. During this slow roll-over,
$\phi_S\equiv\sigma$ is an effectively massless
PNGB field which can act as curvaton.

\section{Curvaton physics}

\paragraph{The required $\varepsilon$}

The phase $\theta=\theta_S\sim 1$ corresponding
to the curvaton degree of freedom remains frozen
until the onset of the curvaton oscillations.
Hence, we expect to have $\sigma_{\rm osc}\sim
\theta v_0$. From \eqref{sosc}, we then find
\begin{equation}
\varepsilon\sim
\frac{\Omega_{\rm dec}}{\pi\zeta\theta}
\left(\frac{m_{3/2}}{M_{\rm P}}
\right)^{\frac{n}{n+1}}\stackrel{_{>}}{_{\sim }}
\varepsilon_{\rm min}\sim
\left(\frac{m_{3/2}}{M_{\rm P}}
\right)^{\frac{n}{n+1}},
\label{epszeta}
\end{equation}
where we have also used that
\mbox{$H_*\sim m_{3/2}$} and
\mbox{$v_0\sim (m_{3/2}M_{\rm P}^n)^{1/(n+1)}$}.
Assuming that $\sigma$ decays before big bang
nucleosynthesis (BBN), i.e. its decay width
$\Gamma_\sigma\sim m_\sigma^3/v_0^2$ is greater
than the Hubble parameter $H_{\rm BBN}$ at
BBN, one shows \cite{amplif,pngbcd} that
\begin{equation}
\varepsilon<
\frac{\Omega_{\rm dec}^{\frac{1}{2}}}{\pi\zeta}
\left(\frac{M_{\rm P}}{T_{\rm BBN}}
\right)^{\frac{1}{2}}
\left(\frac{m_{3/2}}{M_{\rm P}}
\right)^{\frac{5}{4}}\sim
10^{-4}\Omega_{\rm dec}^{\frac{1}{2}},
\label{epsbound0}
\end{equation}
where $T_{\rm BBN}\approx 1~{\rm MeV}$ is the
cosmic temperature at BBN.
From \eqref{epszeta} and \eqref{epsbound0}, we
find \cite{pngbcd} that
\begin{equation}
n>\frac{8+\log(\Omega_{\rm dec}^{\frac{1}{2}}/
\theta)}{7-\log(\Omega_{\rm dec}^{\frac{1}{2}}/
\theta)},
\label{nbound}
\end{equation}
which implies that \mbox{$n\geq 1$} for
\mbox{$\theta\sim 1$}.
The requirement that \mbox{$\Gamma_\sigma>
H_{\rm BBN}$} yields \cite{pngbcd}
\begin{equation}
m_\sigma\stackrel{_{>}}{_{\sim }}
10^{\frac{n-9}{n+1}}~{\rm TeV}\quad\Rightarrow
\quad n\leq\frac{9+\log(m_\sigma/{\rm TeV})}
{1-\log(m_\sigma/{\rm TeV})}
\label{nboundup}
\end{equation}
for \mbox{$m_\sigma<10~{\rm TeV}$}. This
inequality demands that \mbox{$n\leq 9$} for
\mbox{$m_\sigma\stackrel{_{<}}{_{\sim }} 1~
{\rm TeV}$}. So, we get $1\leq n\leq 9$.

\paragraph{Reheating the universe}

We will assume that the curvaton decays after
dominating the energy density of the universe,
i.e. $\Omega_{\rm dec}\simeq 1$, which is
\cite{pngbcd} crucial for avoiding the
overclosure of the universe by axions (see
below). The curvaton dominates \cite{pngbcd}
when \mbox{$H=H_{\rm dom}$}, where
\begin{equation}
H_{\rm dom}\sim
\left(\frac{\sigma_{\rm osc}}{M_{\rm P}}
\right)^4\min\{m_\sigma,~\Gamma_{\rm inf}\}
\label{Hdom}
\end{equation}
with $\Gamma_{\rm inf}\sim g^2m_{3/2}$ being the
inflaton decay width ($g$ is the coupling constant
of the inflaton to its decay products). It can be
shown \cite{pngbcd} that the requirement that
\mbox{$\Gamma_\sigma<H_{\rm dom}$} (i.e. $\sigma$
decays after dominating the energy density of the
universe) results in the bound
\begin{equation}
g>\frac{1}{\theta^2}
\left(\frac{m_{3/2}}{M_{\rm P}}\right)
^{\frac{n-2}{n+1}}\quad\Rightarrow\quad
n>\frac{30-\log\,g-2\log\theta}
{15+\log\,g+2\log\theta}\quad\Rightarrow\quad
n\geq 2\quad{\rm for}\quad\theta\sim 1.
\label{n2}
\end{equation}
Hot big bang begins after the decay of the
curvaton at a reheat temperature
\begin{equation}
T_{\rm REH}\sim\sqrt{\Gamma_\sigma M_{\rm P}}
\sim m_{3/2}\left(\frac{m_{3/2}}{M_{\rm P}}
\right)^{\frac{1}{2}(\frac{n-1}{n+1})}\geq
T_{\rm BBN}\quad{\rm for}\quad n\leq 9.
\label{Treh1}
\end{equation}

\paragraph{Diluting the axions}

For $n>1$, $f_{\rm a}\approx v_0\sim (m_{3/2}
M_{\rm P}^n)^{1/(n+1)}\gg 10^{12}~{\rm GeV}$.
This normally leads to axion overproduction
overclosing the universe. However, if the
curvaton dominates the universe before decaying,
the entropy generated \cite{STurner} during its
decay
\begin{equation}
\frac{S_{\rm after}}
{S_{\rm before}}
\sim\left(\frac{H_{\rm dom}}{\Gamma_\sigma}
\right)^{\frac{1}{2}}
\sim\frac{g\,\theta^2v_0^3}
{m_{3/2}M_{\rm P}^2}\sim
g\theta^2\left(\frac{M_{\rm P}}{m_{3/2}}
\right)^{\frac{n-2}{n+1}}
\label{SS}
\end{equation}
can adequately dilute the axions (see
\cite{lazaetc}).

\paragraph{The evolution of $v$}

The order parameter $v\propto |S|$ must be slowly
rolling during inflation to preserve the
approximate scale invariance of the perturbations
(see below). So, it should follow the equation
\begin{equation}
3H|\dot{S}|+\bar m_S^2|S|\simeq 0\quad\Rightarrow
\quad \frac{\dot v}{v}=
\frac{|\dot S|}{|S|}=\frac{1}{3}c_S
\left(\frac{|m_S^2|}{c_SH^2}-1\right)H.
\label{kg}
\end{equation}
Using \eqref{VN} and the fact that $|m_S^2|\equiv
c_SH^2_{\rm x}\simeq c_SH_{\rm m}^2
(1-e^{-2F_sN_{\rm x}})$, this equation becomes
\begin{equation}
\frac{3}{c_S}\frac{d\ln|S|}{dN}=
\frac{e^{-2F_sN_{\rm x}}-e^{-2F_sN}}
{1-e^{-2F_sN}},
\label{dSN}
\end{equation}
where $H_{\rm x}$ and $N_{\rm x}$ correspond to
the phase transition which changes the sign of
$\bar m_S^2$ during inflation. The solution of
\eqref{dSN} is
\begin{equation}
\frac{6}{c_S}\ln
\left(\frac{|S|_*}{|S|_{\rm x}}\right)
=(1-e^{-2F_sN_{\rm x}})F_s^{-1}\ln
\left(\frac{e^{2F_sN_{\rm x}}-1}
{e^{2F_sN_*}-1}\right)-2(N_{\rm x}-N_*),
\label{SN}
\end{equation}
where \mbox{$|S|_*\equiv |S|(N_*)\sim
(\varepsilon/\varepsilon_{\rm min})H_*$} and
\mbox{$|S|_{\rm x}\equiv |S|(N_{\rm x})\sim
H_{\rm x}/2\pi$} from quantum fluctuations at the
phase transition. The contribution to the
spectral index $n_s$ from the evolution of $v$
during inflation is \cite{pngbcd} $-H_*^{-1}
(\dot{v}/v)_*\leq 0$. The WMAP bound \cite{wmap}
on $n_s$ then implies
\begin{equation}
\frac{c_S}{3}\,e^{-2F_sN_*}\left(
\frac{1-e^{-2F_s(N_{\rm x}-N_*)}}
{1-e^{-2F_sN_*}}\right)\leq 0.04.
\label{nnn}
\end{equation}
It is important to note that, in the present
case, the negative contribution to $n_s$ from
the variation of $v$ during inflation leads
naturally to spectral indices which can be
adequately smaller than unity in accordance with
the recent WMAP results \cite{wmap}.

\section{A concrete example}

\par
From \eqref{nboundup}, and (\ref{n2}) and in view
of \eqref{n4l1}, we see that not many choices for
$n$ are allowed. In fact, we can only accept the
models with \mbox{$n=5,~9$} (i.e. \mbox{$l=1,~2$})
with the latter case being marginal. Hence, to
illustrate the above, we take an example with
\mbox{$n=5$} (i.e. \mbox{$l=1$}) and the
curvaton assuming a random value after the phase
transition, i.e. \mbox{$\theta\sim 1$}.

\par
The bound in \eqref{nboundup} suggests that this
case is acceptable provided that $m_\sigma
\stackrel{_{>}}{_{\sim }}220~{\rm GeV}$. Using
\eqref{epszeta}, we find that $\varepsilon\sim
10^{-8.5}$ (recall that $\Omega_{\rm dec}\simeq
1$) and $\varepsilon_{\rm min}\sim 10^{-12.5}$,
which yields $|S|_*\sim 10^4H_*$. We also
estimate \cite{pngbcd} $N_*$ to be about 38 for
$H_*\sim m_{3/2}$. The reheat temperature turns
out to be $T_{\rm REH}\simeq 10~{\rm MeV}$, while
the entropy production at curvaton decay is given
by $S_{\rm after}/S_{\rm before}\sim 10^{7.5}g$.
From \eqref{n2}, we then conclude that
$g>10^{-4.5}$.

\par
The dilution of axions by the entropy produced
when the curvaton decays after dominating the
universe may lead \cite{choi} to a cosmological
disaster. A sizable fraction of the curvaton's
decay products consists of sparticles, which
eventually turn into stable lightest sparticles
(LSPs) in models (such as ours) with an unbroken
matter parity. The freeze-out temperature of the
LSPs is much higher than $T_{\rm REH}$. Thus, the
LSPs freeze out right after their production and
can, subsequently, overclose the universe. This
problem can be solved \cite{pngbcd} by
suppressing the Higgsino components of the
lighter neutralinos and charginos below $1\%$ and
taking the curvaton adequately light.

\par
One can show \cite{pngbcd} that all the
requirements mentioned above can be satisfied for
\begin{equation}
c_S,~c\stackrel{_{<}}{_{\sim }} O(10^{-4})
\quad\Rightarrow\quad |m_S|,~m_s
\stackrel{_{<}}{_{\sim }} O(10^{-2})\,H_*.
\end{equation}
The smallness of $|m_S|$ is due to the
requirement that $|S|$ is slowly rolling during
the relevant part of inflation so that the
approximate scale invariance of the density
perturbations is preserved, whereas the smallness
of $m_s$ to the required value of $\varepsilon$
($\gg\varepsilon_{\rm min}$), which demands
substantial variation of $|S|$ from the phase
transition during inflation until the time when
the cosmological scales exit the inflationary
horizon. Such a variation can be achieved with a
large number of e-foldings ($N_{\rm x}
\stackrel{_{>}}{_{\sim }}O(10^4)$ and $N_{\rm tot}
\stackrel{_{>}}{_{\sim }}O(10^5-10^6)$), which
means that, in our case, modular inflation is not
of the fast-roll type.

\section{Conclusions}

\par
We constructed SUSY PQ models generating the PQ
scale and the $\mu$ term dynamically. They
contain a successful PNGB curvaton
whose perturbations are suitably amplified to
account for the observed curvature perturbations
even in low-scale inflationary models such as
modular inflation where the inflaton is unable to
generate these perturbations. The spectral
index of density perturbations can easily satisfy
the recent WMAP bound in contrast to other
inflationary models. However, due to the very low
value of the reheat temperature, baryogenesis may
be achieved only via some exotic mechanism (see
\cite{pngbcd}).

\begin{theacknowledgments}
This work was supported by the European Union under
the contract MRTN-CT-2004-503369.
\end{theacknowledgments}

\def\ijmp#1#2#3{{\emph{Int. Jour. Mod. Phys.}}
{\textbf{#1}},~#3~(#2)}
\def\plb#1#2#3{{\emph{Phys. Lett.}}
{\textbf{B~#1}},~#3~(#2)}
\def\zpc#1#2#3{{\emph{Z. Phys.}}
{\textbf{C~#1}},~#3~(#2)}
\def\prl#1#2#3{{\emph{Phys. Rev. Lett.}}
{\textbf{#1}},~#3~(#2)}
\def\rmp#1#2#3{{\emph{Rev. Mod. Phys.}}
{\textbf{#1}},~#3~(#2)}
\def\prep#1#2#3{{\emph{Phys. Rep.}}
{\textbf{#1}},~#3~(#2)}
\def\prd#1#2#3{{\emph{Phys. Rev.}}
{\textbf{D~#1}},~#3~(#2)}
\def\npb#1#2#3{{\emph{Nucl. Phys.}}
{\textbf{B~#1}},~#3~(#2)}
\def\npps#1#2#3{{\emph{Nucl. Phys. B (Proc. Sup.)}}
{\textbf{#1}},~#3~(#2)}
\def\mpl#1#2#3{{\emph{Mod. Phys. Lett.}}
{\textbf{#1}},~#3~(#2)}
\def\arnps#1#2#3{{\emph{Annu. Rev. Nucl. Part. Sci.}}
{\textbf{#1}},~#3~(#2)}
\def\sjnp#1#2#3{{\emph{Sov. J. Nucl. Phys.}}
{\textbf{#1}},~#3~(#2)}
\def\jetp#1#2#3{{\emph{JETP Lett.}}
{\textbf{#1}},~#3~(#2)}
\def\app#1#2#3{{\emph{Acta Phys. Polon.}}
{\textbf{#1}},~#3~(#2)}
\def\rnc#1#2#3{{\emph{Riv. Nuovo Cim.}}
{\textbf{#1}},~#3~(#2)}
\def\ap#1#2#3{{\emph{Ann. Phys.}}
{\textbf{#1}},~#3~(#2)}
\def\ptp#1#2#3{{\emph{Prog. Theor. Phys.}}
{\textbf{#1}},~#3~(#2)}
\def\apjl#1#2#3{{\emph{Astrophys. J. Lett.}}
{\textbf{#1}},~#3~(#2)}
\def\n#1#2#3{{\emph{Nature}}
{\textbf{#1}},~#3~(#2)}
\def\apj#1#2#3{{\emph{Astrophys. J.}}
{\textbf{#1}},~#3~(#2)}
\def\anj#1#2#3{{\emph{Astron. J.}}
{\textbf{#1}},~#3~(#2)}
\def\apjs#1#2#3{{\emph{Astrophys. J. Suppl.}}
{\textbf{#1}},~#3~(#2)}
\def\mnras#1#2#3{{\emph{MNRAS}}
{\textbf{#1}},~#3~(#2)}
\def\grg#1#2#3{{\emph{Gen. Rel. Grav.}}
{\textbf{#1}},~#3~(#2)}
\def\s#1#2#3{{\emph{Science}}
{\textbf{#1}},~#3~(#2)}
\def\baas#1#2#3{{\emph{Bull. Am. Astron. Soc.}}
{\textbf{#1}},~#3~(#2)}
\def\ibid#1#2#3{{\emph{ibid.}}
{\textbf{#1}},~#3~(#2)}
\def\cpc#1#2#3{{\emph{Comput. Phys. Commun.}}
{\textbf{#1}},~#3~(#2)}
\def\astp#1#2#3{{\emph{Astropart. Phys.}}
{\textbf{#1}},~#3~(#2)}
\def\epjc#1#2#3{{\emph{Eur. Phys. J.}}
{\textbf{C~#1}},~#3~(#2)}
\def\nima#1#2#3{{\emph{Nucl. Instrum. Meth.}}
{\textbf{A~#1}},~#3~(#2)}
\def\jhep#1#2#3{{\emph{J. High Energy Phys.}}
{\textbf{#1}},~#3~(#2)}
\def\lnp#1#2#3{{\emph{Lect. Notes Phys.}}
{\textbf{#1}},~#3~(#2)}
\def\appb#1#2#3{{\emph{Acta Phys. Polon.}}
{\textbf{B~#1}},~#3~(#2)}
\def\njp#1#2#3{{\emph{New J. Phys.}}
{\textbf{#1}},~#3~(#2)}
\def\pl#1#2#3{{\emph{Phys. Lett.}}
{\textbf{#1B}},~#3~(#2)}
\def\jcap#1#2#3{{\emph{J. Cosmol. Astropart. Phys.}}
{\textbf{#1}},~#3~(#2)}
\def\mpla#1#2#3{{\emph{Mod. Phys. Lett.}}
{\textbf{A~#1}},~#3~(#2)}

\end{document}